\documentclass[12pt]{article}

\usepackage[left=2.8cm,right=2.8cm,top=2.8cm,bottom=2.8cm]{geometry}

\usepackage[colorlinks=true,linkcolor=black,citecolor=black,urlcolor=blue,filecolor=black]{hyperref}

\usepackage[utf8]{inputenc}
\usepackage{amsmath}
\usepackage{amsfonts}
\usepackage{amssymb}
\usepackage[nosort]{cite}
\usepackage[usenames,table]{xcolor}
\usepackage{dsfont}
\usepackage{mathrsfs}

\usepackage{verbatim}

\newcommand{\be}{\begin{equation}}
\newcommand{\ee}{\end{equation}}

\newcommand{\mfk}[1]{\mathfrak{#1}}

\newcommand{\ihs}{\mfk{ihs}}
\newcommand{\iso}{\mfk{iso}}

\begin{document}

\vspace{30pt}

\begin{center}


{\Large\sc New higher-spin curvatures in flat space\\[10pt]}

\vspace{-5pt}
\par\noindent\rule{350pt}{0.4pt}


\vspace{20pt}
{\sc 
Nicolas Boulanger, Andrea Campoleoni and Simon Pekar
}

\vspace{8pt}
{\it\small
Service de Physique de l'Univers, Champs et Gravitation,\\
Universit{\'e} de Mons -- UMONS,
20 place du Parc, 7000 Mons, Belgium}
\vspace{4pt}

{\tt\small 
\href{mailto:nicolas.boulanger@umons.ac.be}{nicolas.boulanger@umons.ac.be},
\href{mailto:andrea.campoleoni@umons.ac.be}{andrea.campoleoni@umons.ac.be},
\href{mailto:simon.pekar@umons.ac.be}{simon.pekar@umons.ac.be}
}

\vspace{30pt} {\sc\large Abstract} \end{center}

\noindent

It was shown that the Lie algebra underlying higher-spin holography admits a contraction including a
Poincar\'e subalgebra in any space-time dimensions. The associated curvatures, however, do not reproduce
upon linearisation those that are usually employed to formulate the equations of motion of free massless particles
in Minkowski space. We show that, despite this mismatch, the new linearised curvatures can also be used
to describe massless higher-spin fields. This suggests a new way to build interacting higher-spin
gauge theories in Minkowski space that may admit a holographic description.

\newpage



\section{Introduction\label{sec:intro}}

The interactions of massless particles of spin greater than two, 
aka higher-spin particles, are strongly constrained by several no-go results, 
see e.g.~\cite{Bekaert:2010hw} for a review. 
In spite of this, positive results accumulated over the years in an effort motivated,
for instance, by the long-held conjecture that string theory might be a broken phase
of a higher-spin gauge theory and, more recently, by applications in holography.
We refer to \cite{Sagnotti:2011jdy, Giombi:2016ejx} for reviews on these two research
directions and to \cite{Bekaert:2022poo} for a recent status overview of higher-spin theories.
In particular, non-linear equations of motion for massless higher-spin fields on 
constant-curvature backgrounds were built by Vasiliev and collaborators  
\cite{Vasiliev:1990en, Vasiliev:2003ev}. 
Later on, these have been conjectured to provide the bulk duals of 
certain weakly-interacting conformal field theories within the AdS/CFT correspondence \cite{Sezgin:2002rt, Klebanov:2002ja}. 
These developments led to the common lore that higher-spin gauge theories do exist 
in the presence of a cosmological constant~$\Lambda\,$, provided one is ready to 
accept some unconventional features, like e.g.\ an infinite spectrum of fields. 

Vasiliev's equations and higher-spin holography rely upon an infinite-dimensional 
Lie algebra that is essentially unique when the dimension of space-time is bigger 
than three \cite{Fradkin:1986ka, Boulanger:2013zza}, modulo supersymmetric 
extensions and Chan-Paton factors \cite{Konstein:1989ij, Vasiliev:2004cm, Sezgin:2012ag}.
For instance, Vasiliev's equations are built using curvatures valued in 
this higher-spin algebra that we shall denote by $\mathfrak{hs}_D\,$, 
with $D$ the space-time dimension. 
This approach to the interactions of fields of arbitrary spin,
often referred to as unfolded formulation, somehow extends to higher spins
the Cartan formulation of general relativity. 
In one of the founding papers of the unfolded formulation, it was observed that 
the algebra $\mathfrak{hs}_4$ admits a contraction containing a Poincar\'e 
subalgebra \cite{Fradkin:1986ka}. 
The result, however, was discarded as a candidate higher-spin algebra in Minkowski
space because the associated linearised curvatures for spin $s>2$ 
do not agree with those emerging from the first-order free action of \cite{Vasiliev:1980as}.
Equivalently, while the linearised curvatures of fields with $s=2$ agree 
with the usual torsion and Riemann curvature, for $s \geqslant 3$ they
do not agree with the $\Lambda \to 0$ limit of those entering the
equations of motion for free massless particles on (Anti) de Sitter ((A)dS)
space of \cite{Lopatin:1987hz}.

This observation was long considered as an additional 
no-go argument against higher-spin interactions in Minkowski space: 
no appropriate symmetry algebra seemed to exist, at least for the 
same spectrum of fields as in Vasiliev's equations. 
This view was also supported by direct analyses of interactions 
within Fronsdal's metric-like formulation \cite{Fronsdal:1978rb} 
in which a particle of spin $s$ is described starting from a 
rank-$s$ symmetric tensor.
In this setup, various studies independently pointed out 
the inconsistency of the non-Abelian, two-derivative, 
minimal gravitational coupling of Fronsdal's gauge fields in flat space;
see, e.g., \cite{Bekaert:2010hw, Joung:2013nma}
and references therein. 
As discussed in \cite{Bekaert:2010hw}, Weinberg's famous 
low-energy theorem \cite{Weinberg:1964} as well as the generalised 
Weinberg-Witten theorem of \cite{Porrati:2008rm} can also be reinterpreted 
in these terms. 

On the other hand, in \cite{Boulanger:2008tg},  
a consistent non-Abelian cubic coupling between massless spin-$s$ 
and spin-$2$ fields around Minkowski space 
containing a total of \mbox{$(2s-2)$} derivatives 
was obtained in Fronsdal's formulation. 
Although it was shown to induce a consistent non-Abelian 
deformation of the free gauge algebra satisfying Jacobi 
identities \cite{Bekaert:2010hp}, its analysis has not been pushed 
to next (quartic and higher) orders in the fields, which would be needed to complete an interacting theory. 
More recently, a complete interacting higher-spin gauge theory on 
four-dimensional flat manifolds with Euclidean or split signature, 
dubbed chiral higher-spin gravity, has been built employing
a different set of fields \cite{Ponomarev:2016lrm, Krasnov:2021nsq}.
The role of chiral higher-spin models in flat-space holography 
also begins to be explored \cite{Ren:2022sws, Ponomarev:2022ryp, Monteiro:2022xwq}. 
Besides, an analogue of the contraction of the algebra $\mathfrak{hs}_4$ 
discussed in \cite{Fradkin:1986ka} was recently defined in any space-time 
dimension \cite{Campoleoni:2021blr}. The contracted algebra, that we 
shall denote as $\mathfrak{ihs}_D$, can also be obtained from the 
Poincar\'e algebra following a construction close to that relating 
its AdS ancestor $\mathfrak{hs}_D$ to the conformal 
algebra \cite{Campoleoni:2021blr, Bekaert:2022oeh}.

These indications naturally lead to reconsider the linearised curvatures 
of $\mathfrak{ihs}_D\,$. 
In this note, we propose a new system of first-order equations of motion 
built upon them, that describes the free propagation of massless particles 
of arbitrary spin on Minkowski space. 
Our equations follow the same pattern as in the usual unfolded 
formulation: for any spin $s$ we set to zero all corresponding 
curvatures but one, and impose that the 
latter is proportional to the spin-$s$ Weyl tensor. 
We then prove that, even if our curvatures have a non-standard form, 
the resulting equations of motion are equivalent to the standard 
ones \cite{Lopatin:1987hz}.

When expressed in terms of curvatures, the structure of our 
equations is the same as that of the customary free unfolded equations in AdS.
Moreover, the full non-linear curvatures are a contraction of the AdS ones. 
This strongly suggests the option to deform our linear equations into an 
interacting theory following the path that led from \cite{Fradkin:1986ka} 
to \cite{Vasiliev:1990en, Vasiliev:2003ev} or, equivalently, the cohomological 
approach of \cite{Sharapov:2017yde, Sharapov:2019vyd}.
We defer a detailed analysis to future work, but we wish to stress that
this programme is expected to provide, on the one hand, a 
model for higher-spin interactions in Minkowski space-time and,
on the other hand, a simple and concrete model for flat-space holography
(see, e.g., \cite{deBoer:2003vf, Arcioni:2003td, Ciambelli:2018wre, Pasterski:2021raf, Donnay:2022aba, Bagchi:2022emh} for an overview of various approaches).
The higher-spin algebra $\mathfrak{ihs}_D$ indeed also appears as a subalgebra
of the higher symmetries of a Carrollian scalar living on null infinity 
\cite{Bekaert:2022oeh}. 
This is the analogue of a pillar of higher-spin holography:
the symmetry algebra $\mathfrak{hs}_D$ is realised at the boundary of AdS
as the algebra of higher symmetries of a free conformal scalar 
\cite{Nikitin1991-gp, Eastwood:2002su}.
All higher-spin dualities involve deformations of this basic setup
preserving such symmetry, see e.g.~\cite{Bekaert:2012ux} for a review 
focusing on this aspect.
Any non-linear deformation of our free equations of motion will
therefore provide a candidate gravitational dual of the simplest
Carrollian field theory, thus fitting within the urgent quest
for concrete dual pairs in flat-space holography, that is
currently mainly driven by symmetry considerations.

\section{Higher-spin extension of the Poincar\'e algebra} \label{sec:algebra}

The higher-spin algebra $\ihs_D$ can be obtained as
an \.In\"on\"u-Wigner contraction of the algebra $\mathfrak{hs}_D$
underlying Vasiliev's equations in AdS space \cite{Campoleoni:2021blr}. 
The latter can be built, e.g., by evaluating the universal enveloping algebra
of $\mfk{so}(2,D-1)$ on Dirac's singleton module \cite{Eastwood:2002su, Iazeolla:2008ix, Boulanger:2011se, Joung:2014qya}. 
Its generators can be collected in irreducible and traceless
tensors $M_{a(s),b(t)}$ with $s \geqslant 0$ and $0 \leqslant t \leqslant s$,
where the shorthand $a(s)$ denotes a set of $s$ symmetrised indices. 
They thus satisfy $\eta^{cd} M_{a(s-2)cd,b(t)} = 0$ and $M_{a(s),ab(t-1)} = 0$, 
where repeated indices denote a symmetrisation with strength one,
and correspond to representations of the Lorentz algebra
labelled by two-row Young tableaux. Their commutators take the form
\begin{equation} \label{grading}
\left[ M_{a(s_1),b(t_1)} , M_{c(s_2),d(t_2)} \right] \propto \sum_{s_3 = |s_1-s_2|+1}^{s_1+s_2-1} \sum_{t_3=0}^{s_3} M_{e(s_3),\,f(t_3)}
\end{equation}
with $(s_1+s_2+s_3) \mod 2 = 1$ and $(t_1+t_2+t_3) \mod 2 = 1$.
For $s = 1$ one recovers the $\mfk{so}(2,D-1)$ conformal algebra
and explicit structure constants can be found in \cite{Joung:2014qya}.

The generators $M_{a(s),b(t)}$ with $s-t$ even thus form a subalgebra
and one can rescale the others as
\begin{equation} \label{contraction}
M_{a(s),b(s-2n-1)} \to \epsilon^{-1} M_{a(s),b(s-2n-1)} \, , \quad \forall\ s,\, n \in \mathbb{N} . 
\end{equation}
In the limit $\epsilon \to 0$ all commutators involving only generators
with $s-t$ odd vanish, while the others remain untouched.
The $\mathfrak{so}(2,D-1)$ subalgebra contracts into a $\iso(1,D-1)$ subalgebra
and one obtains a non-Abelian higher-spin extension of the Poincar\'e algebra.
We chose to present this algebra as a contraction of the AdS higher-spin algebra
$\mathfrak{hs}_D$, although one can also build it as a quotient of
the universal enveloping algebra of the Poincar\'e algebra \mbox{$\iso(1,D-1)$} \cite{Campoleoni:2021blr, Bekaert:2022oeh}.
One can also prove that, in a generic space-time dimension $D > 3$,
$\mathfrak{ihs}_D$ is the only algebra with the same set of generators as
$\mathfrak{hs}_D$ that can be built with this procedure \cite{Campoleoni:2021blr}.
When $D=3$, the option to build sensible higher-spin algebras in Minkowski space
from contractions of the AdS ones was already observed in \cite{Blencowe:1988gj} (see also \cite{Afshar:2013vka, Gonzalez:2013oaa, Ammon:2017vwt, Campoleoni:2021blr}),
and Vasiliev-like equations of motion were proposed in \cite{Ammon:2020fxs}.

\section{New equations of motion in Minkowski space}

We now consider a one-form taking values in the Lie-algebra $\ihs_D$,
\begin{equation}
A = \sum_{s=0}^\infty \sum_{t=0}^{s}\omega^{a(s),b(t)} M_{a(s),b(t)} \, ,
\end{equation}
and its Yang-Mills curvature,
\begin{equation}
{\rm d}A + A \wedge A = \sum_{s=0}^\infty \sum_{t=0}^{s} F^{a(s),b(t)} M_{a(s),b(t)} \, .
\end{equation}
The one-forms $\omega^a$ and $\omega^{a,b}$ correspond, respectively, to 
the space-time vielbein and spin connection.

We wish to build equations of motion describing free massless particles 
using the linearisation of the curvatures $F^{a(s),b(t)}$ around the 
Minkowski background, linearisation that we shall denote as $\bar{F}^{a(s),b(t)}\,$. 
We thus split the vielbein as $\omega^a = h^a + e^a$ and, for simplicity, 
we choose Cartesian coordinates so that the background vielbein reads 
$h_\mu{}^a = \delta_\mu{}^a$ and the background spin connection vanishes. 
The following discussion can be extended to arbitrary coordinates by 
introducing a flat background Lorentz connection, but working in Cartesian 
coordinates makes some arguments more transparent. 

The linearised curvatures only depend on the commutators between the 
higher-spin generators $M_{a(s),b(t)}$ and those of the Poincar\'e 
subalgebra. With our choice of coordinates, they read
\begin{subequations} \label{curvatures}
\begin{alignat}{5}
\bar{F}^{a(s),b(t)} & = {\rm d} \omega^{a(s),b(t)} \quad & & \textrm{for $s-t$ even} , \label{curv-even} \\[5pt]
\bar{F}^{a(s),b(t)} & = {\rm d} \omega^{a(s),b(t)} + h^{\{b} \wedge \omega^{a(s),b(t-1)\}} + h_c \wedge \omega^{a(s),b(t)c} \quad & & \textrm{for $s-t$ odd} , \label{curv-odd}
\end{alignat}
\end{subequations}
where braces denote a two-row Young projection together with a traceless 
projection, so that the corresponding term shares the same symmetry properties as the others: 
\begin{equation}
\begin{split}
&(s-t+2)\, h^{\{b} \wedge \omega^{a(s),b(t-1)\}} := (s-t+1)\, h^b \wedge \omega^{a(s),b(t-1)} - s \, h^a \wedge \omega^{a(s-1)b,b(t-1)} \\
& \quad - \tfrac{s}{D+s+t-4}\, \eta^{ab} \left( (s-t)\, h_c \wedge \omega^{a(s-1)c,b(t-1)} - 
(t-1)\tfrac{D+2s-4}{D+2t-6}\, h_c \wedge \omega^{a(s-1)b,cb(t-2)} \right) \\
& \quad + \tfrac{s(s-1)}{D+s+t-4}\, \eta^{aa} \left( h_c \wedge \omega^{a(s-2)cb,b(t-1)} - \tfrac{t-1}{D+2t-6}\, h_c \wedge \omega^{a(s-2)b(2),cb(t-2)} \right) \\
& \quad + \tfrac{(t-1)(s-t+1)}{D+2t-6}\, \eta^{bb} h_c \wedge \omega^{a(s),b(t-2)c} \,,
\end{split}
\end{equation}
where we recall that repeated indices denote a symmetrisation with strength one,
e.g.\ $A_a B_a := \frac{1}{2} \left( A_{a_1} B_{a_2} + A_{a_2} B_{a_1} \right)$.
Notice that the second term on the r.h.s.\ of \eqref{curv-odd} is absent for $t = 0$,
when this value is allowed by the parity condition. The curvature with $t=s$,
instead, always fits in the class \eqref{curv-even}.
The linearised curvatures \eqref{curvatures} are invariant under the gauge transformations
\begin{subequations} \label{gauge}
\begin{alignat}{5}
\delta \omega^{a(s),b(t)} & = {\rm d} \lambda^{a(s),b(t)} \quad & & \textrm{for $s-t$ even} , \label{gauge-even} \\[5pt]
\delta \omega^{a(s),b(t)} & = {\rm d} \lambda^{a(s),b(t)} + h^{\{b}\,\lambda^{a(s),b(t-1)\}} + h_c\,\lambda^{a(s),b(t)c} \quad & & \textrm{for $s-t$ odd} . \label{gauge-odd}
\end{alignat}
\end{subequations}

To describe a particle with spin $s$ we propose to impose the equations of motion 
\begin{subequations} \label{eoms}
\begin{align}
\bar{F}^{a(s-1),b(t)} & = 0 \,, \qquad  0 \leqslant t \leqslant s-2 \,, \label{eom1}\\[5pt]
\bar{F}^{a(s-1),b(s-1)} & = h_c \wedge h_d\, C^{a(s-1)c,b(s-1)d} \, , \label{eom2}
\end{align}
\end{subequations}
where $C^{a(s),b(s)}$ is a gauge-invariant and Lorentz-irreducible tensor. 
The integrability of eq.~\eqref{eom2} also imposes a tower of Bianchi identities on this zero-form, leading to its interpretation as Weyl tensor, see e.g.~\cite{Bekaert:2004qos, Didenko:2014dwa}.
For $s=2$, eqs.~\eqref{eoms} are the linearised vacuum Einstein equations,
where the vanishing of the Ricci tensor is reformulated by equating
the Riemann curvature with the Weyl tensor.
In the following, we prove that they describe the free propagation of
a massless particle of spin $s$ by showing that they are equivalent to
the Lopatin-Vasiliev equations of motion on Minkowski space \cite{Lopatin:1987hz}; see also \cite{Skvortsov:2008vs, Skvortsov:2008sh, Boulanger:2008up}. 

Alternatively, one can obtain eqs.~\eqref{eoms} by rescaling 
$\omega^{a(s-1),b(s-2n)} \to \epsilon\, \omega^{a(s-1),b(s-2n)}$ in the equations
of motion of \cite{Lopatin:1987hz} and sending $\epsilon \to 0$ while keeping their
dependence on the cosmological constant $\Lambda$ fixed. The latter can then
be absorbed in a redefinition of the connections $\omega^{a(s-1),b(s-2n-1)}\,$.
If one instead keeps $\epsilon$ fixed while sending $\Lambda \to 0$, one obtains
a different limit, displayed in eqs.~\eqref{LV_general} below, which is
what we refer to as the Lopatin-Vasiliev equations on Minkowski space.  

\section{Equivalence with the Lopatin-Vasiliev equations}

To prove that eqs.~\eqref{eoms} are equivalent to the Lopatin-Vasiliev equations
\cite{Lopatin:1987hz} and, therefore, to the Fronsdal equation in Minkowski space
\cite{Fronsdal:1978rb}, we begin with the instructive spin-three example.
We then extend the proof to any spin.

\subsection{The spin-three example}

For a spin-three particle eqs.~\eqref{eoms} read
\begin{subequations} \label{spin3-eom}
\begin{align}
\bar{T}^{ab} & := {\rm d} e^{ab} = 0 \, , \label{eom3-1} \\
\bar{T}^{ab,c} & := {\rm d} \omega^{ab,c} + h^{\{c} \wedge e^{ab\}} + h_d \wedge X^{ab,cd} = 0 \, , \label{eom3-2} \\
\bar{R}^{ab,cd} & := {\rm d} X^{ab,cd} = h_e \wedge h_f\, C^{abe,cdf} \, , \label{eom3-3}
\end{align}
\end{subequations}
where, for clarity, we renamed the fields $\omega^{ab} \to e^{ab}$ and $\omega^{ab,cd} \to X^{ab,cd}$ and we stressed that $\bar{F}^{ab}$ and $\bar{F}^{ab,c}$ play the role of torsions, while $\bar{F}^{ab,cd}$ plays that of a curvature. These equations are invariant under
\begin{subequations} \label{spin3-gauge}
\begin{align}
\delta e^{ab} & = {\rm d} \xi^{ab} \,, \label{gauge3-1} \\
\delta \omega^{ab,c} & = {\rm d} \lambda^{ab,c} + h^{\{c}\, \xi^{ab\}} + h_d\, \rho^{ab,cd} \,, \label{gauge3-2} \\
\delta X^{ab,cd} & = {\rm d} \rho^{ab,cd} \,. \label{gauge3-3}
\end{align}
\end{subequations}

Thanks to the Poincar\'e lemma, eq.~\eqref{eom3-1} implies that $e^{ab}$ is pure gauge. We can thus set it to zero using the gauge symmetry generated by $\xi^{ab}$. In this gauge, eqs.~\eqref{eom3-2} and \eqref{eom3-3} take the same form as the ($\Lambda \to 0$ limit) of the corresponding Lopatin-Vasiliev equations.
To show that these two equations suffice to describe a massless spin-three particle,
it is enough to notice that, in the gauge $e^{ab} = 0$, eq.~\eqref{eom3-2} implies
\begin{equation}
h_c \wedge \bar{T}^{ab,c} = h_c \wedge {\rm d} \omega^{ab,c} = - \, {\rm d} \left( h_c \wedge \omega^{ab,c} \right) = 0 \, .
\end{equation}
This is the case because $X^{ab,cd}$ is symmetric in its last two indices and with our choice for the background vielbein, ${\rm d} h^a = 0$. 
The Poincar\'e lemma then allows one to introduce the one-form $\tilde{e}^{ab}$ such that
\begin{equation}
- h_c \wedge \omega^{ab,c} = {\rm d} \tilde{e}^{ab} \,.
\end{equation}
This relation is valid for all $\omega^{ab,c}$, in particular for 
a pure-gauge infinitesimal configuration $\delta\omega^{ab,c}$
for which we denote the corresponding r.h.s.\ of the previous equation by ${\rm d} \delta\tilde{e}^{ab}\,$.
The configuration $\delta\tilde{e}^{ab}$ is then seen to be identically equal to 
\begin{equation}
\delta \tilde{e}^{ab} = {\rm d} \tilde{\xi}^{ab} + h_c\,\lambda^{ab,c}
\end{equation}
(recall that we used $\xi^{ab}$ to fix the gauge $e^{ab} = 0$, so that $\tilde{\xi}^{ab}$ is a new gauge parameter that does not affect $\omega^{ab,c}$).

The fields $\tilde{e}^{ab}$, $\omega^{ab,c}$ and $X^{ab,cd}$ manifestly satisfy the $\Lambda \to 0$ limit of the Lopatin-Vasiliev equations, that is 
\begin{subequations} \label{LV}
\begin{align}
{\rm d} \tilde{e}^{ab} + h_c \wedge \omega^{ab,c} & = 0 \, , \label{LV3-1} \\
{\rm d} \omega^{ab,c} + h_d \wedge X^{ab,cd} & = 0 \, , \label{LV3-2} \\
{\rm d} X^{ab,cd} & = h_e \wedge h_f\, C^{abe,cdf} \, , \label{LV3-3}
\end{align}
\end{subequations}
with the gauge symmetries
\begin{subequations}
\begin{align}
\delta \tilde{e}^{ab} & = {\rm d} \tilde{\xi}^{ab} + h_c\,\lambda^{ab,c} \,, \\
\delta \omega^{ab,c} & = {\rm d} \lambda^{ab,c} + h_d\, \rho^{ab,cd} \,, \\
\delta X^{ab,cd} & = {\rm d} \rho^{ab,cd} \,.
\end{align}
\end{subequations}

For the reader's convenience, we recall that once the form 
\eqref{LV} of the equations of motion is reached, 
one can use the parameter $\lambda^{ab,c}$ to gauge away the corresponding component of 
$h^{\mu\,c}\,\tilde{e}_{\mu}{}^{ab}\,$, 
so as to recover the Fronsdal field 
$\varphi_{\mu\nu\rho} =  h_{(\mu}{}^{a} h_{\nu}{}^{b} \tilde{e}_{\rho)\,ab}\,$. 
The torsion constraint \eqref{LV3-1} then allows one to express $\omega^{ab,c}$ in terms of 
the first derivative of $\varphi_{\mu\nu\rho}\,$, 
except for a pure-gauge component which is gauged away using $\rho^{ab,cd}\,$. 
The torsion constraint \eqref{LV3-2} plays a double role: some of its irreducible components only involve $\omega^{ab,c}$ and impose Fronsdal's equation on $\varphi_{\mu\nu\rho}\,$, 
while the others express $X^{ab,cd}$ in terms of 
the first derivatives of $\omega^{ab,c}$ and, eventually, in terms of two derivatives of 
the Fronsdal field. 
The last equation \eqref{LV3-3} 
then expresses the Weyl tensor in terms of third derivatives of $\varphi_{\mu\nu\rho}$. For a more detailed review of this mechanism see, e.g., \cite{Skvortsov:2008sh}.

\subsection{Arbitrary spin}

For an arbitrary value $s$ of the spin, the equations $\bar{F}^{a(s-1),b(t)} = 0$ with $s-t-1$ even and greater than zero imply that the fields $\omega^{a(s-1),b(t)}$ are pure gauge thanks to the Poincar\'e lemma. As such, they can be set to zero using the gauge variations \eqref{gauge-even}:
\begin{equation}
\bar{F}^{a(s-1),b(s-2n-1)} = 0 \quad \Rightarrow \quad \omega^{a(s-1),b(s-2n-1)} 
\xrightarrow[\textrm{gauge fixing}]{} 0 \,, \quad \textrm{for}\ 1 \leqslant n \leqslant \left\lfloor\frac{s-1}{2} \right\rfloor .
\end{equation}
In this gauge, most of the other torsion-like equations become closure conditions too:
\begin{equation}
\bar{F}^{a(s-1),b(s-2n)} = 0 \quad \xrightarrow[\textrm{gauge fixing}]{} \quad {\rm d} \omega^{a(s-1),b(s-2n)} = 0 \,, \quad \textrm{for}\ 2 \leqslant n \leqslant \left\lfloor\frac{s}{2} \right\rfloor .
\end{equation}
This shows that the fields $\omega^{a(s-1),b(s-2n)}$ with $n \geqslant 2$ can also be eliminated using the parameters $\lambda^{a(s-1),b(s-2n)}$ with $n \geqslant 2$. Eventually, one is left with the equations
\begin{subequations} \label{final-eom}
\begin{align} 
\bar{F}^{a(s-1),b(s-2)} & := {\rm d} \omega^{a(s-1),b(s-2)} + h_c \wedge \omega^{a(s-1),b(s-2)c} = 0 \,, \label{final1} \\[5pt]
\bar{F}^{a(s-1),b(s-1)} & := {\rm d} \omega^{a(s-1),b(s-1)} = h_c \wedge h_d\, C^{a(s-1)c,b(s-1)d} \,. \label{final2}
\end{align}
\end{subequations}
The first one implies
\begin{equation}
h_c \wedge \bar{F}^{a(s-1),b(s-3)c} = - \, {\rm d} \left( h_c \wedge \omega^{a(s-1),b(s-3)c} \right) = 0
\end{equation}
and, thanks to the Poincar\'e lemma,
\begin{equation}
- h_c \wedge \omega^{a(s-1),b(s-3)c} = {\rm d} \tilde{\omega}^{a(s-1),b(s-3)} \,.
\end{equation}
The procedure can be iterated to obtain
\begin{equation}
- {\rm d} \left(h_c \wedge \tilde{\omega}^{a(s-1),b(s-k)c} \right) = 0 \quad \Rightarrow \quad - h_c \wedge \tilde \omega^{a(s-1),b(s-k)c} = {\rm d} \tilde{\omega}^{a(s-1),b(s-k)} \,,
\end{equation}
for all $4 \leqslant k \leqslant s$, thus reconstructing the full Lopatin-Vasiliev system of equations in Minkowski space:
\begin{subequations} \label{LV_general}
\begin{align} 
{\rm d} \tilde{\omega}^{a(s-1),b(t)} + h_c \wedge \tilde{\omega}^{a(s-1),b(t)c} & = 0 \,, \qquad 0 \leqslant t \leqslant s-2 \,, \\[5pt]
{\rm d} \tilde{\omega}^{a(s-1),b(s-1)} & = h_c \wedge h_d\, C^{a(s-1)c,b(s-1)d} \,, 
\end{align}
\end{subequations}
where it is understood that $\tilde{\omega}^{a(s-1),b(s-1)} := \omega^{a(s-1),b(s-1)}$ and $\tilde{\omega}^{a(s-1),b(s-2)} := \omega^{a(s-1),b(s-2)}$.
These equations are then invariant under
\begin{equation}
\delta \tilde{\omega}^{a(s),b(t)} = {\rm d} \tilde{\lambda}^{a(s),b(t)} + h_c\,\tilde{\lambda}^{a(s),b(t)c} \, ,
\end{equation}
where $\tilde{\lambda}^{a(s-1),b(s-1)}$ and $\tilde{\lambda}^{a(s-1),b(s-2)}$ coincide with the gauge parameters with the same structure in eqs.~\eqref{gauge}.

The same argument can be repeated using generic coordinates on Minkowski 
space-time: the main modification is that the exterior derivative has to be traded for the background Lorentz-covariant derivative $\nabla$. This does not affect the previous analysis because we only used that the latter annihilates the background vielbein, $\nabla h^a = 0$, and it is nilpotent.

\section{Discussion}

We proposed new first-order equations of motion for free massless particles 
of arbitrary spin in Minkowski space, built upon the linearised curvatures
of the flat-space higher-spin algebra $\mathfrak{ihs}_D$ introduced 
in \cite{Campoleoni:2021blr, Bekaert:2022oeh}.
If the set of fields that we use in our equations is the same as in 
Lopatin-Vasiliev's equations on (A)dS background \cite{Lopatin:1987hz}, 
that are the free equations of motion on top of which Vasiliev's unfolded 
formulation for interacting higher-spin fields is constructed, 
we stress that their precise expressions differ from the 
zero cosmological constant limit of the equations in \cite{Lopatin:1987hz}.
In spite of this difference, that a priori could prevent one 
from eliminating some auxiliary fields, 
we showed that our equations nevertheless propagate the correct 
degrees of freedom for a massless field in Minkowski space-time of dimension 
$D \geqslant 4$.
This is so because all fields $\omega^{a(s-1),b(t)}$ with $t \leqslant s-3$ are 
actually pure gauge, while the field equations involving $\omega^{a(s-1),b(s-1)}$, 
which encode the degrees of freedom via the Weyl tensor,
take the same form in both systems of equations.

As a result, the non-linear curvatures of the non-Abelian,
flat-space higher-spin algebra $\mathfrak{ihs}_D$  
of \cite{Campoleoni:2021blr} can be considered as the basic 
building blocks to construct an interacting higher-spin gauge 
theory in Minkowski space in the unfolded formalism, 
along the purely algebraic lines of \cite{Vasiliev:1990en, Vasiliev:2003ev} 
or of its reformulation in \cite{Sharapov:2017yde, Sharapov:2019vyd}.
The fact that the flat-space higher-spin algebra $\mathfrak{ihs}_D$ 
possesses an Abelian ideal, contrary to the AdS algebra $\mathfrak{hs}_D$, 
suggests even more freedom in introducing interactions 
via the cohomological approach of \cite{Sharapov:2017yde, Sharapov:2019vyd}.

Another remark supporting the proposal to build 
a non-linear theory based on the algebra $\mathfrak{ihs}_D$ 
is that such a theory should include the  $(2s-2)$-derivative 
coupling of a massless spin-$s$ field to gravity of
\cite{Metsaev:2005ar, Boulanger:2008tg}.
Indeed, in \cite{Boulanger:2008tg} it was shown that this cubic vertex induces a non-Abelian deformation of the free gauge algebra, 
leading to a contribution proportional to the translation generator 
$P_a$ in the commutator $\left[M^{a(s-1),b(s-1)},M^{c(s-1),d(s-2)}\right]\,$.
As explained in Section \ref{sec:algebra}, the latter commutator 
is unaffected by the contraction leading from $\mfk{hs}_D$ to $\mfk{ihs}_D\,$,
which does contain a contribution proportional to $P_a\,$.
Moreover, as discussed in \cite{Boulanger:2008tg}, the $(2s-2)$-derivative 
vertex is the one that possesses the highest number of derivatives among those 
that constitute the Fradkin-Vasiliev gravitational coupling in AdS 
\cite{Fradkin:1987ks},
that has later been reproduced within the unfolded formulation.
Only this top vertex survives the flat limit  
that coincides with the high-energy limit of the Fradkin-Vasiliev 
action, thereby evading the low-energy no-go results 
\cite{Weinberg:1964,Porrati:2008rm}.

Finally, let us stress that our result suggests the option to define an 
interacting dual of a Carrollian scalar on null infinity, 
with similar features to the bulk models 
entering higher-spin holography \cite{Giombi:2016ejx}. The Carrollian approach to flat-space holography, see 
e.g.~\cite{Donnay:2022aba, Bagchi:2022emh}, appears particularly well suited 
to this proposal, which relies on the observation that a Carrollian scalar admits a $\mathfrak{ihs}_D$ symmetry algebra \cite{Bekaert:2022oeh};
see also \cite{Grumiller:2019tyl} for similar bulk/boundary realisations of higher-spin symmetries in $D=3$.

\section*{Acknowledgments}

We thank E.~Skvortsov for discussions. 
A.C.\ and S.P.\ are, respectively, a research associate and a FRIA grantee of the Fonds de la Recherche Scientifique~--~FNRS.
This work was partially supported by FNRS under Grants No.\ FC.36447, F.4503.20 and T.0022.19.
S.P.\ also acknowledges the support of the SofinaBo\"el Fund for Education and Talent. A.C.\ and S.P.\ thank the \'Ecole polytechnique and S.P.\ thanks the School of Mathematics and the Maxwell Institute for Mathematical Sciences of the University of Edinburgh for hospitality.



\end{document}